\documentclass[prl, aps, twocolumn, floatfix, superscriptaddress, amsmath,  longbibliography]{revtex4-1}
\usepackage{graphicx}
\usepackage{graphics}
\usepackage{mathtools}
\usepackage{float}
\usepackage{amssymb}
\usepackage{bbold}

\usepackage[dvipsnames]{xcolor}
\definecolor{darkblue}{rgb}{0.0, 0.0, 0.75}

\usepackage{color}
\usepackage[colorlinks=true,
            linkcolor=darkblue,
            urlcolor=darkblue,
            citecolor=darkblue]{hyperref}

	\usepackage[normalem]{ulem} 
	\definecolor{mgreen}{RGB}{1,123,0}

\def \br{{\bf r}}

\def \zb{{\bar{z} }}

\def \mb{\text{b}}
\def \mJ{\text{J}}
\def \ms{\text{s}}

\def \tw{\tilde{\omega} }

\def \tV{\tilde{V} }

\def \mp{\mathrm{p}}

\def \mext{\mathrm{ext}}

\def \mB{\mathrm{B}}
\def \ms{\mathrm{s}}
\def \mm{\mathrm{m}}

\def \mum{\mu\mathrm{m} }

\def \cD{\mathcal{D}}

\def \mHz{\mathrm{Hz}}

\def \mms{\mathrm{ms}}

\def \cD{{\cal{D}}}

\def \tV{\tilde{V} }

\begin{document}
\title{Shapiro steps in driven atomic Josephson junctions}
\author{Vijay Pal Singh}
\affiliation{Quantum Research Centre, Technology Innovation Institute, Abu Dhabi, UAE}
%
\author{Juan Polo}
\affiliation{Quantum Research Centre, Technology Innovation Institute, Abu Dhabi, UAE}
\author{Ludwig Mathey}
\affiliation{Zentrum f\"ur Optische Quantentechnologien and Institut f\"ur  Quantenphysik, Universit\"at Hamburg, 22761 Hamburg, Germany}
\affiliation{The Hamburg Centre for Ultrafast Imaging, Luruper Chaussee 149, Hamburg 22761, Germany}
\author{Luigi Amico}
\thanks{On leave from the Dipartimento di Fisica e Astronomia ``Ettore Majorana'', University of Catania.}
\affiliation{Quantum Research Centre, Technology Innovation Institute, Abu Dhabi, UAE}
\affiliation{INFN-Sezione di Catania, Via S. Sofia 64, 95127 Catania, Italy}
\affiliation{Centre for Quantum Technologies, National University of Singapore 117543, Singapore}
\date{\today}
%


%
\begin{abstract}
We study driven atomic Josephson junctions realized  by coupling two two-dimensional atomic clouds with a tunneling barrier. 
By moving the barrier at a constant velocity, dc and  ac Josephson regimes  are characterized by a zero and nonzero atomic density difference across the junction,  respectively. 
Here, we monitor the dynamics resulting in the  system when, in addition to the above constant velocity protocol,  the position of the barrier is  periodically  driven.  We demonstrate that the time-averaged particle imbalance features a plateau behavior that is the analog of Shapiro steps observed in driven superconducting Josephson junctions.  
The underlying dynamics reveals an intriguing interplay of the vortex and phonon excitations, 
where Shapiro steps are induced via suppression of vortex growth. We study the system with a classical-field dynamics method, and benchmark our findings with a driven circuit  dynamics.  
\end{abstract}
\maketitle
%
%

%
\begin{figure}
\includegraphics[width=1.0\linewidth]{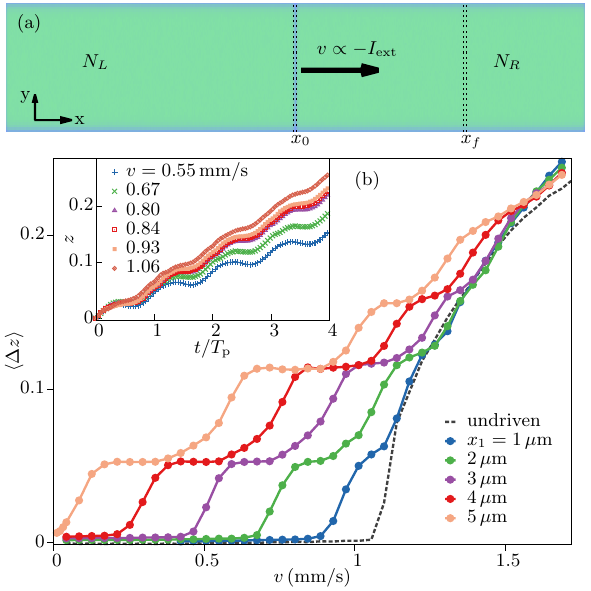}
\caption{Atomic Josephson junction and emergence of Shapiro steps.
(a) Simulation of a Josephson junction consisting of two clouds separated by a tunneling barrier of height $V_0$ 
and width $w$. The barrier is moved at constant velocity $v$ (arrow) from the position $x_0$ until the final position $x_f$, 
which induces an external current $I_\mext$. 
In addition, we modulate the barrier location using the protocol $x(t) = vt + x_1 \sin(2\pi f t)$,  
where $f$ is the frequency and $x_1$ is the amplitude of driving. 
$N_L$ ($N_R$) represents the atom number of the left (right) reservoir. 
(b) Imbalance $\Delta z = z- \zb$ as a function of $v$ for the undriven case (dashed line).  
$\zb$ is the equilibrium imbalance determined at $x_f$ that varies with $v$.
Time-averaged imbalance $\langle \Delta z \rangle$ shown for $f=45\, \mHz$ and $x_1= 1, 2, 3, 4$ and $5\, \mum$. 
Inset shows the time evolution of $z(t)$ at different $v$ for the driving with $x_1=3\, \mum$, 
where $T_\mp =1/f$ is the driving period.  
We use the barrier height $V_0/\mu=1.5$, where $\mu$ is the mean-field energy. 
}
\label{Fig:system}
\end{figure}

Superconducting Josephson junctions (JJs) exhibit a transition between the dc and ac Josephson effect by developing a dc voltage  
when the current exceeds a critical value  \cite{Josephson1962}. 
In the presence of microwave radiation driving the junction, characteristic dc-ac transitions can occur as a result of photon-assisted tunneling processes. Accordingly, for an averaged voltage matching multiples of the  driving frequency,  the  supercurrent  jumps  between different `dc plateaux', reflecting that the  Cooper pairs phase change is effectively synchronized with  the external ac source \cite{GrossNotes}. The resulting steps displayed in the I-V characteristic are referred to as Shapiro steps \cite{Shapiro1963, Grimes1968}. 
Such picture  has been confirmed with experiments carried out on driven superconducting JJs  \cite{Hebboul1990,Rosenbach2021,Yan2023}. Shapiro steps play an important role both for the fundamental understanding of superconductivity, and practical applications of JJ, such as metrological voltage standards \cite{Hamilton1995,Burroughs1999,Burroughs2011}.

Similarly to the dc and ac Josephson effects in superconducting junctions,
dissipationless-viscous  transitions can occur also in neutral $^3$He and $^4$He quantum fluids flowing through suitable constrictions.
This phenomenon has been demonstrated to result  from   phase slips nucleating in the hydrodynamical field \cite{Avenel1988,Hoskinson2006,Backhaus1997, Davis2002}. 
By suitable driving of the pressure across the constriction, a matter-flow with Shapiro-type step behaviour was  reported in superfluid $^3$He \cite{Simmonds2001}.

   Ultracold atoms have emerged as ideal systems to implement and study atomtronic analogues of superconducting circuits  
\cite{Amico2017, Amico2021, Amico2022, Ramanathan2011, Eckel2014, Ryu2013,  Ryu2020, Chien2015, Krinner2017}. 
Atomic Josephson junctions (JJs) were realized using weak links of atom clouds \cite{Cataliotti2001, LeBlanc2011, Spagnolli2017,  Pigneur2018}, enabling the study of important effects as macroscopic self-trapping \cite{Albiez2005} and  current-phase relation \cite{Luick2020, Kwon2020}. 
 Following  the seminal paper of Giovannazzi et al \cite{Giovanazzi2000}, the  Josephson effect in ultracold atom systems can be studied by  moving  a barrier separating two degenerate gas clouds  at rest. An analog of a  dc to ac Josephson  response occurs, in which the difference of the particles densities of the two sides of the junction (termed as particle imbalance throughout the manuscript)  changes from zero to a finite value.  The current-chemical potential (playing the role of the voltage in superconducting JJ) relation was achieved in a series of remarkable experiments \cite{Levy2007, Kwon2020, Pace2021}. 
  Such a behaviour can be captured by a resistively and capacitively shunted junction (RCSJ) circuit model \cite{ch11}.  
Phase-slip induced dissipation was studied in Refs. \cite{Burchianti2018, Singh2020jj, Xhani2020, Xhani2022, Gabriel2023}. 
Theoretically, Josephson effects were studied using a  two-mode Gross-Pitaevskii equation (GPE) model \cite{Smerzi1997,Raghavan1999,Giovanazzi2000}. 
Even though, specific resonances were noticed in periodically tilted double wells, experimentally feasible protocols to realize Shapiro steps in condensates have not been provided yet \cite{Raghavan1999, Kohler2003, Eckardt2005, Grond2011}.

With the remarkable progress in dynamical light shaping achieved recently \cite{Ryu2015,rubinsztein2016roadmap,barredo2018synthetic},  driven matter-wave circuits are well within the experimental capabilities.
Here, we consider a driven JJ: While  moving  the barrier with a velocity $v$, we  drive its position periodically - see Fig. \ref{Fig:system}(a). 
With this, we observe that the dynamics of the system features steps of the  Shapiro type that manifest themselves as a time-averaged atom imbalance - see Fig. \ref{Fig:system}(b). We demonstrate how the observed dynamics is controlled by  specific features of  phonon  excitations and vortex nucleation. 
The results are obtained by classical field methods that include fluctuating bosonic fields beyond mean-field description \cite{Blakie2008, Polkovnikov2010,  Singh2016, Singh2020sound}. 
We benchmark our results with the dynamics of a  driven RCSJ circuit  model.

\begin{figure*}
\includegraphics[width=1.0\linewidth]{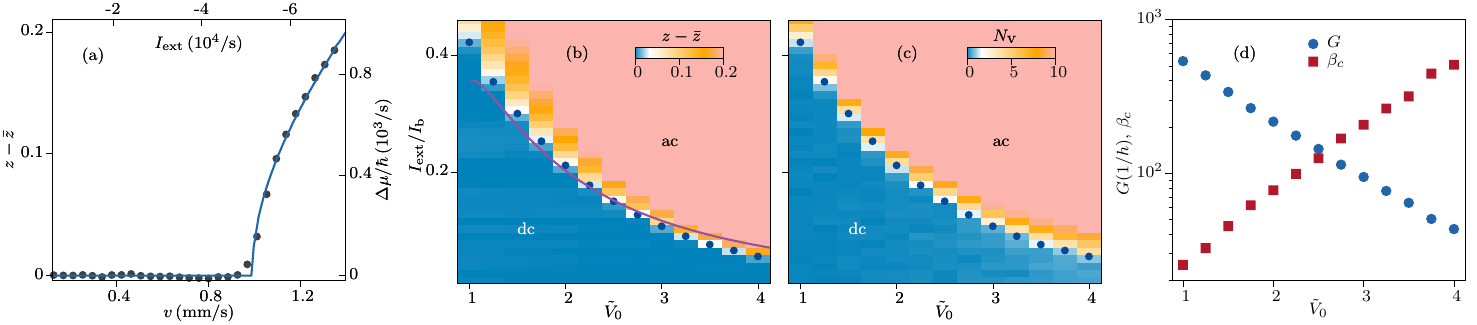}
\caption{Characterizing an atomic Josephson junction.
(a) Imbalance  $z- \zb$ as a function of $v$ for $\tV_0=1.5$ (dots) and its correspondence to the current-chemical potential results (second axes).  
We fit the response with $\langle \Delta \mu \rangle=  \sqrt{I_\mext^2 - I_c^2}/G$ (continuous line) to determine the critical current $I_c$ and the conductance $G$. 
(b) $z- \zb$ as a function of $\tV_0$ and $I_\mext/I_\mb$, where $I_\mb$ is the bulk current. 
The results of $I_c$ (dots) are compared with the theoretical prediction $I_{c, \mp}$ (continuous line); see text.
(c) Total vortex number $N_\text{v}$ determined from the same simulations as in (b), which shows a sharp onset of vortex growth above $I_c$ (dots).  
(d) Conductance $G$ (dots) and Stewart-McCumber parameter $\beta_c$ (squares) determined for $\tV_0$ in the range $1-4$.   }
\label{Fig:dc}
\end{figure*}

{\it{System and method.}---} 
We simulate the dynamics of a generic condensate of bosons using classical-field dynamics within the truncated Wigner approximation \cite{Blakie2008, Polkovnikov2010, Singh2016, Singh2020sound}.  
We consider a homogeneous cloud of bosons confined in a box of dimensions $L_x \times L_y$. 
The system is described by the Hamiltonian
\begin{equation} \label{eq_hamil}
\hat{H}_{0} = \int \mathrm{d}{\bf r} \Big[ \frac{\hbar^2}{2m} \nabla \hat{\psi}^\dagger({\bf r}) \cdot \nabla \hat{\psi}({\bf r})  + \frac{g}{2} \hat{\psi}^\dagger({\bf r})\hat{\psi}^\dagger({\bf r})\hat{\psi}({\bf r})\hat{\psi}({\bf r})\Big].
\end{equation}
$\hat{\psi}$ ($\hat{\psi}^\dagger$) is the bosonic annihilation (creation) operator. 
The interaction $g=\tilde{g} \hbar^2/m$ is given in terms of the dimensionless parameter $\tilde{g} =  \sqrt{8 \pi} a_s/\ell_z$, 
where $m$ is the mass, $a_s$ is the s-wave scattering length and $\ell_z= \sqrt{\hbar/(m \omega_z)}$ is the harmonic oscillator length in the transverse direction. 
Within the classical-field representation we replace the operators $\hat{\psi}$ in Eq. (\ref{eq_hamil}) and the equations of motion by complex numbers $\psi$.  
The initial states $\psi (\br,t=0)$ are sampled in a grand canonical ensemble with chemical potential $\mu$ and temperature $T$ via a classical Metropolis algorithm \footnote{See Supplemental Material for the simulation method, the two-mode GPE equations, the estimate of the critical current, the ac dissipation mechanism, the scaling of characteristic parameters and the time-averaged vortex number, which includes Refs. \cite{Giovanazzi2000,Meier2001, Kohler2003, Singh2020jj,Ambegaokar1963,Kwon2020}.}.
The resulting distribution provides the fluctuations of $\psi (\br,t=0)$ around its mean field value. 
Finally, each initial state is propagated using the equations of motion
\begin{align}\label{eq:eom}
 i \hbar \dot{\psi}(\br, t) = \Bigl(  - \frac{\hbar^2}{2m} \nabla^2 + V(\br, t) + g|\psi|^2 \Bigr) \psi(\br, t),
\end{align}
which include the barrier potential given by $V({\bf r},t)  = V_0 (t) \exp \bigl[- 2 \bigl( x-x(t) \bigr) ^2/w^2 \bigr]$. $V_0(t)$, $w$  and $x(t)$ are the barrier's  strength, width and location.  
For numerical calculations, we discretize space on a lattice of size $N_x \times N_y$ and a discretization length $l=0.5\, \mu \mathrm{m}$.   
While we present our results for the concrete realization provided by $^{6}\mathrm{Li}_2$ molecules, we emphasize that our protocol can be applied to any cold-atom degenerate gas. 
We choose the density $n \approx 5.6\, \mu \mm^{-2}$, $\tilde{g}=0.1$,  $T/T_0=0.06$ and $L_x \times L_y = 512 \times 27\, \mum^2$.  
The critical temperature $T_0$ is estimated by $T_0 = 2\pi n \hbar^2/(m k_\mB \cD_c)$, 
where $\cD_c= \ln(380/\tilde{g})$ is the critical phase-space density \cite{Prokofev2001, Prokofev2002}. 
We use $w/\xi = 1.1$ and $V_0$ in the range $ V_0/\mu \equiv \tilde{V}_0 = 1-4$, 
where $\xi= \hbar/\sqrt{2m gn}$ is the healing length and $\mu = gn$ is the mean-field energy. 
To create the weak link at the location $x(t)= x_0= L_x/2$, we ramp up $V_0$ linearly over $200\, \mms$ and wait for $50\, \mms$. 
Following  \cite{Giovanazzi2000,Kwon2020}, the Josephson current can be obtained by moving the barrier at a constant velocity $v$ until it reaches the final position $x_f$, 
as depicted in Fig. \ref{Fig:system}(a). 
Here, we will discuss the dynamics of the system obtained when the barrier features a  periodic driving in addition to its motion with constant velocity: 
\begin{align} \label{eq:protocol}
x(t) = vt + x_1 \sin(2\pi f t), 
\end{align}
where $x_1$ is the amplitude and $f$ is the frequency of driving.  
We calculate the atom number $N_L(t)$ ($N_R(t)$) in the left (right) reservoir to determine the imbalance $z(t)= (N_L(t) - N_R(t))/N$, where $N=(N_L + N_R)$ is the total atom number.  
The barrier motion induces an external current given by $I_\mext (t) = (\zb N/2) \times |v|/\Delta x$, where $\Delta x$ is the displacement and $\zb$ is the equilibrium imbalance at the final location $x_f = x_0 + \Delta x$.
Throughout the paper we fix the driving time $t_f$ to $4$ cycles.  
We calculate $z(t)$ for various values of $v$, see inset of Fig. \ref{Fig:system}(b). 
By fitting $z(t)$ with a linear function we obtain the value of the time-averaged imbalance $\langle \Delta z \rangle$ at $t_f$, where $\langle \Delta z \rangle = \langle z \rangle - \zb(x_f)$ with $x_f = |v| \times t_f$. 
$\langle ..\rangle$ refers to the time-averaged response throughout the paper.
In Fig. \ref{Fig:system}(b),  the driven response shows the formation of Shapiro steps in comparison to the undriven system, which we explain below.  
For the undriven case, we show $ \Delta z= z- \zb(x_f)$ determined at the same $t_f$ as the driven case. 
We quantify the change in the chemical potential $\Delta \mu =  N E_c \Delta z/2$, where $E_c = 4 (\partial \mu/\partial N)$ is the effective charging energy.

{\it{Characterization of the dc-ac regimes.}---} 
We first introduce the driven RCSJ  model and then analyze the dc-ac regimes of the  undriven junction. 
We note that this is intrinsically different from the two-mode GPE equations, which converge to the RCSJ model in the ideal underdamped limit only. 
The RCSJ model is  a lumped elements circuit to  model the dynamics of JJs \cite{ch11}.  
The Kirchhoff law of the driven RCSJ circuit  reads
\begin{align}
I_\mext  + I_1 \tw \cos(\omega t) &= I_c \sin \phi - G \Delta \mu - C \Delta \dot{\mu},  \label{eq:rcsj-I} 
\end{align}
where $I_c$ is the critical current, $\phi= \phi_L - \phi_R$  the phase difference across the junction, $G$ the conductance and $C=1/E_c$ the capacitance. The Josephson relation for the phase dynamics is 
$
\hbar \dot{\phi} = - \Delta \mu$, meaning that $\Delta \mu$ plays the role of the voltage  across the junction.
$I_1 \tw $ is the amplitude and $\omega$ is the frequency of ac drive, with $\tw = \omega/\omega_\mJ$, where $\omega_\mJ = \sqrt{I_c E_c/\hbar} $ is the Josephson frequency.
By using the expression of $\Delta \mu$ we provided above,  the circuit is described as an effective driven resistively shunted junction model 
\begin{align}\label{eq:rcsj2}
\dot{z}N /2 + I_1  \tw \cos(\omega t) &= I_c \sin \phi + \hbar G  \dot{\phi}. 
\end{align}
The undriven case (i.e., $I_1=0$) is solved analytically, yielding the time-averaged chemical potential $\langle \Delta \mu \rangle = G^{-1} \sqrt{I_\mext^2 - I_c^2}$, 
which we use to fit our results for the $\Delta \mu(I_{\mext},t_f)$ curve. 
Based on this fit, we determine  $I_c$ and $G$. 
Here, for the undriven case, we use the constant displacement of $\Delta x= 150\, \mum$, which results in the same equilibrium imbalance $\zb = 0.59$ at $x_f$ for all $v$.
As shown in Fig. \ref{Fig:dc}(a), there is a nonzero $\Delta \mu$ only when $I_\mext$ exceeds $I_c$,  
which marks the transiton from the dc to ac Josephson effect.
The junction becomes resistive with a finite $\Delta \mu$ above $I_c$. 
We map out the ac resistive regime for a wide range of $\tV_0$ in Fig. \ref{Fig:dc}(b).
The onset of the resistive regime occurs at a low value of $I_c$ for high $\tV_0$. 
We compare the results of $I_c$ with the predictions of the critical current $I_{c,\mp} = I_\mb t_0 (\tV_0, w)$ derived for an ideal Josephson junction \cite{Singh2020jj}. 
The bulk current $I_\mb = c n_0 L_y$ is determined using the sound velocity $c=\sqrt{gn/m}$ and the condensate density $n_0$. 
We use the variational solution of the tunneling amplitude $t_0 (\tV, d)$ obtained across a rectangular barrier of width $d$ and height $\tV$ \cite{Singh2020jj}, with $d \approx 1.1 w$ and 
$\tV = \tV_0$ \cite{Note1}.
In Fig. \ref{Fig:dc}(b) the results of $I_{c,\mp}$ show good agreement with the simulation results.

To characterize the resistive regime, we study  the dynamics of the condensate's local density and phase.  
While we observe no  distinctive change in the dynamics for $I_\mext$ below $I_c$, characteristic density patterns due to phonon and vortex excitations, resulting in an increase of the imbalance, occur for $I_\mext$ above $I_c$ \cite{Note1}.  
We identify vortex excitations by calculating the phase winding around the lattice plaquette of size $l\times l$ using $\sum_{\Box} \delta \theta(x,y) = \delta_x \theta(x,y) + \delta_y\theta(x+l,y)+\delta_x\theta(x+l,y+l)+\delta_y\theta(x,y+l)$, 
where $ \theta(x,y) $ is the phase field of $\psi(x,y)$ and  the phase differences between sites are taken to be $\delta_{x/y} \theta(x,y)  \in (-\pi, \pi]$. 
We associate a vortex (an antivortex) by a phase winding of $2\pi$ ($-2\pi$). 
By counting all vortices and antivortices we determine the total vortex number $N_\text{v}$ and average it over the initial ensemble.  
Fig. \ref{Fig:dc}(c) shows a rapid growth of $N_\text{v}$ above $I_c$ and no vortex excitation below $I_c$. 
This confirms that the energy in the resistive regime is dissipated by the creation of vortex-antivortex pairs, 
which is analogous to phase-slips in $^4$He  \cite{Avenel1988, Hoskinson2006} and atomic weak links  \cite{Burchianti2018}.

 In Fig. \ref{Fig:dc}(d), we demonstrate that the dependence of both the conductance $G$ and the Stewart-McCumber parameter $\beta_c =  I_c C/(\hbar G^2)$ is consistent with an exponential dependence on $\tV_0$. 
The characteristic dependence between $G$ and $I_c$ yields to $G \propto I_c^{\alpha}$, with $\alpha=1.2$ \cite{Note1}, 
resembling the linear dependence of the  Ambegaokar-Baratoff relation \cite{Ambegaokar1963}. 
The values of $\beta_c$ are in the range between $25$ and $500$, which fulfills the junction dynamics being in the underdamped regime 
($\beta_c \gg 1$). The dependence of  $\beta_c$ on $G$ follows the behavior  $\beta_c \propto G^{-1.17}$ \cite{Note1}.

\begin{figure}
\includegraphics[width=1.0\linewidth]{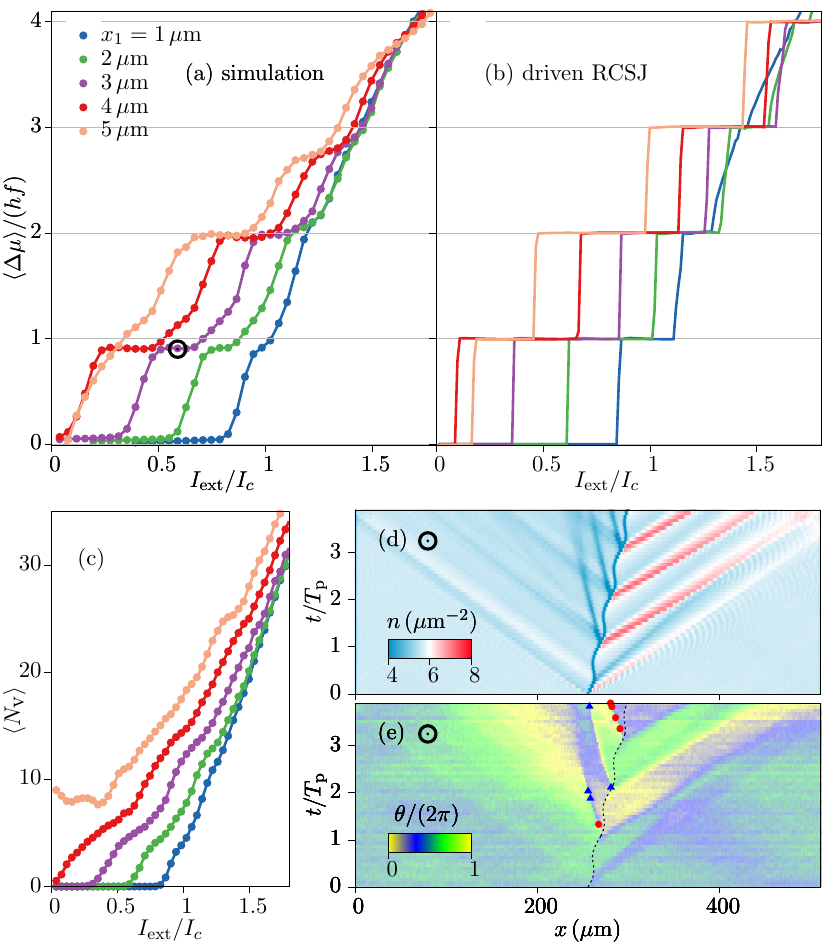}
\caption{Shapiro steps and the underlying vortex and phonon dynamics. 
(a) $\langle \Delta \mu \rangle$--$I_\mext$ response of the driven junction for $f/f_\mJ= 2.2$ and $x_1= 1, 2, 3, 4$ and $5\, \mum$. The corresponding undriven junction with $\tV_0= 1.5$ gives $I_c = 1.6 \times 10^5\, \ms^{-1}$, $\hbar G= 157$ and the Josephson frequency $f_\mJ=25 \, \mHz$. 
(b) Results of the driven RCSJ model shown for the same parameters as the simulations. 
(c) Time-averaged vortex number $\langle N_\text{v} \rangle$ corresponding to the simulations in (a). 
(d, e)  Time evolution of the density $n(x)$ and the phase $\theta (x, L_y/2)$ in the $x$ direction of the driven system at $I_\mext/I_c=0.6$, 
which is indicated by the open circle symbol in (a). 
$T_\mp =1/f$ is the driving period.
$n(x)$ is averaged over the initial ensemble and the $y$ direction, 
whereas $\theta (x, L_y/2)$ represents the phase profile of a single sample. 
In (e), vortices (triangles) and antivortices (dots) are calculated using the phase change near the line at $L_y/2$ and the dotted line denotes the barrier motion. 
 }
\label{Fig:step}
\end{figure}

\begin{figure}
\includegraphics[width=1.0\linewidth]{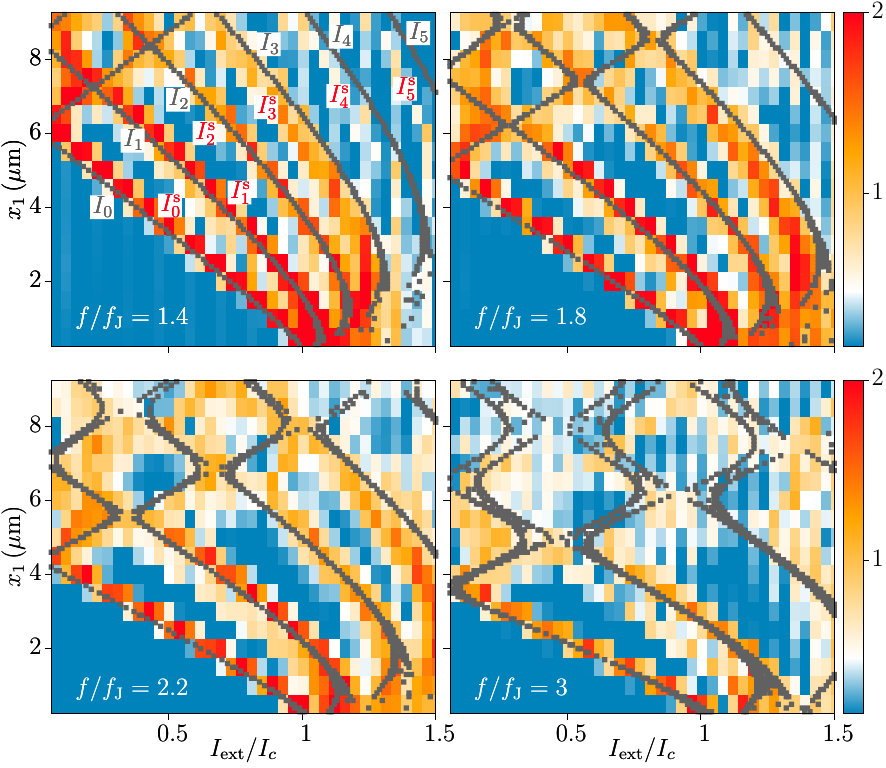}
\caption{Maximal current at the Shapiro steps. 
Differential resistance $d\mu /d I_\mext$ as a function of $I_\mext/I_c$ and $x_1$, for $f/f_\mJ=1.4$, $1.8$, $2.2$ and $3$. 
The maxima of $d\mu /d I_\mext$ allow us to identify the maximal currents $I_0^\ms$, $I_1^\ms$, $I_2^\ms$, $I_3^\ms$, $I_4^\ms$, and $I_5^\ms$ of the Shapiro steps $0$, $1$, $2$, $3$, $4$ and $5$, respectively. 
The grey squares show the frequency and amplitude dependence of the maximal currents $I_k$ of the driven RCSJ model, determined using the same magnitude cutoff for all $f$, where the splitting of curve indicates a broadening of maxima. 
$k$ is the step index. We use $I_c = 1.6 \times 10^5\, \ms^{-1}$, $f_\mJ=25 \, \mHz$, and $\hbar G$ in the range  $145 -175$, which are chosen according to the undriven system. }
\label{Fig:max}
\end{figure}

{\it{Driven response and Shapiro steps.}---} 
We now turn to the dynamics of a periodically driven junction, which is obtained using the barrier protocol described in Eq. \ref{eq:protocol}, see also Fig. \ref{Fig:system}(a).  
In Fig. \ref{Fig:step}(a) the time-averaged response of $\langle \Delta \mu \rangle$ features the creation of regular steps occurring at $\langle \Delta \mu \rangle= k h f$,  
where $k$ is an integer denoting the step index. 
The onset current location and width of the step vary according to the value of $x_1$.  
To benchmark our results we numerically solve the driven RCSJ model and analyze its time-averaged response for various parameters, 
where the current-driving amplitude $I_1$ is related to $x_1$ via $x_1 = I_1 v_c/(I_c \omega_\mJ)$ 
in which $v_c$ is the critical velocity associated with $I_c$ \cite{Note1}. 
The results of the driven model confirm the formation of the Shapiro steps at $\langle \Delta \mu \rangle = k h f$ in agreement with our simulations, see Fig. \ref{Fig:step}(b).
The steps obtained by the driven RCSJ model are smoothed out due to the thermal fluctuations of the initial state and a driving-induced depletion of the condensate density. 
As a result, especially for $k>2$, the steps are obtained  with a specific broadening and amplitude suppression.

Below, we monitor the formation of vortices taking place while the system is driven. 
Fig. \ref{Fig:step}(c) shows the total number $\langle N_\text{v} \rangle $ of vortices  nucleating during driving, 
averaged over $2$ driving cycles \cite{Note1}.  
Remarkably, $\langle N_\text{v} \rangle $ features a similar plateau structure as the $\langle\Delta \mu \rangle$ response: 
$\langle N_\text{v} \rangle $ increases rapidly at the onset of the  steps and  undergoes a suppressed  vortex nucleation during the steps formation.
For $x_1= 5\, \mum$, $\langle N_\text{v} \rangle $ shows a distinct behavior compared to the results at lower $x_1$. 
$\langle N_\text{v} \rangle$ starts out with a nonzero value undergoing a plateau structure during the first step, whose onset and width feature an opposite trend than the corresponding results of low $x_1$ in Figs. \ref{Fig:step}(a) and (b). 
This occurs due to a Bessel-function type feature of the step width on $x_1$, 
which is evident from the results shown in Fig. \ref{Fig:max}.
Even though  $I_\mext$ is below $I_c$,  as a consequence of the periodic driving,  both density wave pulses in $n(x,y)$ and  slippages  of the local phase  $\theta(x,y)$  near the barrier occur  at the beginning of each driving cycle. 
Vortex formation and density waves are clearly correlated:  vortex-antivortex pairs are generated which  
propagate behind the barrier; at the same time, density wave pulses of low velocity are observed in the left reservoir. 
 Such phenomenon occurs in correspondence of the maxima,  where the current near the barrier exceeds $I_c$, making the barrier effectively dissipative \cite{Note1}, see Figs. \ref{Fig:step}(d, e).

 To quantify the maximal current that develops at each step, we monitor the maxima $I_k^\ms$ of the differential resistance $d\mu/d I_\mext$ at different  driving frequencies. 
Fig. \ref{Fig:max} shows that the numerically obtained $I_k^\ms$ follow the behaviour of maximal currents of the driven RCSJ  model. 
The amplitude of the peak decreases for both high steps and high driving frequencies, where the latter is partially captured by the driven model. The high-step damping originates from the depletion of the condensate density, which is a feature of the dynamics and does not affect the circuit model.

{\it{Conclusions and outlook.}---}
We have analyzed the time-averaged response of a driven atomic Josephson junction (JJ) at nonzero temperature. 
The JJ is created by separating two two-dimensional bosonic clouds with a tunneling barrier, 
where the barrier motion induces an external current and the periodic modulation of the barrier position acts as an external ac current drive. For the analysis, we employed   classical-field simulations that capture the dynamics beyond mean field.
The driven response demonstrates  dc-ac transitions in the form of Shapiro steps. 
We compare these results with a driven RCSJ circuit model. 
Indeed, the steps result from resonances between the driving frequency $f$ and the periodic oscillations in the particle imbalance such that $\langle\Delta \mu \rangle = k h f$, 
where $k$ is the step index.
As a distinctive feature of our neutral superfluid system, the phenomenon arises from characteristic dynamics of vortex and phonon excitations. 
Our results can be directly probed, for example on ultracold $^6$Li machines employed in LENS \cite{Kwon2020} and Hamburg \cite{Luick2020}.  
Because of the possibility to tune interactions from negative to positive values and relying on the know-how of the field  allowing to work with bosonic and/or fermionic systems, spinor condensates \cite{Evrard2019} etc, 
Shapiro steps are expected to bear a great potential to explore the coherent properties of the artificial quantum matter as provided by cold atoms.   
Our results are important for both fundamental research in quantum dynamics of coherent systems and applications in quantum technologies.

{\it{Acknowledgments.}---}  We thank Giulia Del Pace, Giampiero Marchegiani and Giacomo Roati for  discussions. 
We acknowledge PHYSnet computing resources of University of Hamburg. 
L. M. acknowledges funding by the Deutsche Forschungsgemeinschaft (DFG) in the framework of SFB 925 – project ID 170620586 and the excellence cluster  `Advanced Imaging of Matter’ - EXC 2056 - project ID 390715994.

\bibliography{References}

\end{document}


\title{Supplemental Material for ``Shapiro steps in driven atomic Josephson junctions''}
\author{Vijay Pal Singh}
\affiliation{Quantum Research Centre, Technology Innovation Institute, Abu Dhabi, UAE}
%
\author{Juan Polo}
\affiliation{Quantum Research Centre, Technology Innovation Institute, Abu Dhabi, UAE}
\author{Ludwig Mathey}
\affiliation{Zentrum f\"ur Optische Quantentechnologien and Institut f\"ur  Quantenphysik, Universit\"at Hamburg, 22761 Hamburg, Germany}
\affiliation{The Hamburg Centre for Ultrafast Imaging, Luruper Chaussee 149, Hamburg 22761, Germany}
\author{Luigi Amico}
\thanks{On leave from the Dipartimento di Fisica e Astronomia ``Ettore Majorana'', University of Catania.}
\affiliation{Quantum Research Centre, Technology Innovation Institute, Abu Dhabi, UAE}
\affiliation{INFN-Sezione di Catania, Via S. Sofia 64, 95127 Catania, Italy}
\affiliation{Centre for Quantum Technologies, National University of Singapore 117543, Singapore}
\date{\today}
%
%
%
\maketitle
%
%

%
%
%
%

\section{Classical-field dynamical method}
In this section, we elaborate on our simulation method. 
We sample the initial states $\psi (\br,t=0)$ in a grand canonical ensemble with temperature $T$ and chemical potential $\mu$ via a classical Metropolis algorithm. 
We fix $T/T_0=0.06$, while the value of  $\mu$ is set to obtain an average density of $n \approx 5.6\, \mu \mm^{-2}$, 
where $T_0$ is an estimate of the critical temperature. 
The simulation procedure is the following. 
We select a lattice site randomly and perform single-site updates by changing real and imaginary parts of the wave function, 
which are chosen from a normal distribution. The width of the distribution is adjusted such that the acceptance rate of each step is around one half. We perform about $10^5$ steps to obtain the thermalized state. 
After that we generate the initial states $\psi (\br,t=0)$ of the system by performing more than $5000$ updates per site, so that the generated states are uncorrelated. Each initial state is propagated using the equations of motion to obtain the time evolution $\psi (\br,t)$.

\section{Two-mode Gross-Pitaevskii equation}
Here we outline the two-mode Gross-Pitaevskii equation (GPE) model, which was considered to study Josephson effects in atomic condensates \cite{Giovanazzi2000}. 
We start with the GPE in one dimension
\begin{align}\label{eq:GPE}
i\hbar \frac{\partial}{\partial t} \psi(x, t)  = \bigl[ - \frac{\hbar^2}{2m} \frac{\partial^2}{\partial x^2} + V_\text{ext} + g |\psi|^2 \bigr] \psi (x, t),
\end{align}
where $\psi(x, t)$ is the condensate field, $g$ is the interaction and $V_\text{ext}(x, t) = V_0 e^{-2(x-vt)^2/w^2}$ is the barrier potential. $v$ is the barrier velocity. Under the approximation of weak coupling between the left and right condensates we use the ansatz \cite{Giovanazzi2000}
\begin{align}
\psi(x, t) = \psi_1(t) \Phi_1 (x)  +  \psi_2(t) \Phi_2 (x).
\end{align}
$ \psi_{n}(t) = \sqrt{N_{n} (t)} \exp[ - i \theta_{n} (t) ]$ are the time-dependent complex amplitudes of the left ($n=1$) and right condensate ($n=2$). 
 $\Phi_{1,2} (x)$  are orthonormal and approximate ground state solutions of the GPE of the left and right condensates. 
The time evolution of the imbalance $z = (N_1 -N_2)/N$ and the phase difference $\theta = \theta_1 -  \theta_1$ is governed by the equations \cite{Giovanazzi2000}
\begin{align}
\hbar \dot{ z } &= (2E_J/N) \sqrt{1- z^2} \sin(\theta),  \label{eq:eta}\\
\hbar \dot{ \theta}  &= \frac{\hbar \omega_\mp^2  v t}{v_c} - \frac{2E_J}{N} \frac{z}{\sqrt{1- z^2} } \cos(\theta) -  \frac{NE_c}{2} z    \label{eq:theta}, 
\end{align}
where $E_J$ is the coupling energy across the barrier, $E_c=2 g \int dx |\Phi_1(x)|^4$ is the capacitive energy, and $\omega_\mp = \sqrt{E_J E_c}/\hbar$ is the Josephson frequency. We use $\tau = \omega_\mp t$ and obtain
\begin{align}
\frac{\partial z }{ \partial \tau}  &=  \alpha^{-1}  \sqrt{1- z^2 } \sin(\theta),   \label{eq:eta1} \\
\frac{\partial \theta }{ \partial \tau}  &=  ( \tilde{v} \tau -  \alpha z) - \frac{ \alpha^{-1} z}{ \sqrt{1-z^2} } \cos(\theta) \label{eq:theta1},
\end{align}
where $\alpha = (N/2) \sqrt{E_c/E_J}$ and $\tilde{v} = v/v_c$. 
Eqs. \ref{eq:eta1} and \ref{eq:theta1} describe the dynamics of a moving barrier 
separating two condensates, which was shown to undergo the dc-ac transition \cite{Giovanazzi2000}. 
Since there are no dissipative terms, the dynamics in long time evolution suffers from capacitive effects \cite{Meier2001}, 
which are important for quantitative description as in  our case.
We use the transformation $\tilde{z} =  ( \tilde{v} \tau -  \alpha z) $, yielding
\begin{align}
\dot{ \tilde{z} } &=  \tilde{v}  - \sqrt{1- (\tilde{v} \tau - \tilde{z}  )^2/\alpha^2 } \sin(\theta),  \\
 \dot{ \theta}  &=  \tilde{z}  - \frac{ (\tilde{v} \tau - \tilde{z}  ) }{ \alpha^2 \sqrt{1- (\tilde{v} \tau - \tilde{z}  )^2/\alpha^2} } \cos(\theta).
\end{align}
For large $\alpha$ we then find
\begin{align}
\dot{ \tilde{z} } & \approx  \tilde{v}  -  \sin(\theta),  \\
 \dot{ \theta}  & \approx  \tilde{z}.
\end{align}
This results in 
\begin{align}\label{eq:gpec}
\tilde{v}=  \frac{d^2 \theta}{d \tau^2}  + \sin(\theta). 
\end{align}
We add an ac driving of the barrier location using  $x(t) = vt + x_1 \sin(\omega t)$ in Eq. \ref{eq:theta1}. 
Eq. \ref{eq:gpec} takes the form 
\begin{align}\label{eq:bec2nd}
\tilde{v} + \tilde{v}_1 \tilde{\omega} \cos (\tilde{\omega} \tau) &=  \frac{d^2 \theta}{d \tau^2}  + \sin(\theta) 
\end{align}
where $\tilde{v}_1 =x_1 \omega_\mp/v_c $ and $\tilde{\omega} = \omega/\omega_\mp$. 
The equivalence $I_1/I_c = x_1 \omega_\mp/v_c $ gives $I_1 =x_1 \omega_\mp I_c/v_c  $.

The dynamics described by Eqs. \ref{eq:eta1} and \ref{eq:theta1} is distinct from that of Josephson junctions in the RCSJ circuit model. 
This is namely twofold.  
First, there is a nonlinear cosine term in Eq. \ref{eq:theta1}, resulting in an additional coupling of the phase and current as compared to  the RCSJ model.  
Second, the two-mode GPE model does not account for a dissipative current, which is important for stable junction dynamics. 
Here, phenomenological-type of a damping term can be added to address the existence of a  dissipative current \cite{Kohler2003}. Though, the two-mode GPE model is instructive, it is dominant by capacitive effects. As a consequence of this,  the two-mode GPE model approaches the RCSJ circuit model  only in the ideal underdamped limit  (i.e. without conductance), see Eq. \ref{eq:gpec}.


%
\begin{figure}
\includegraphics[width=1.0\linewidth]{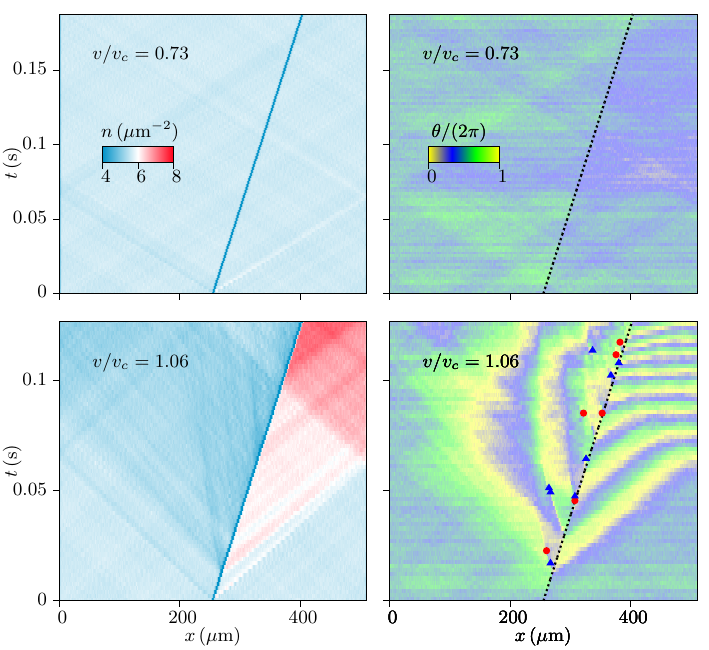}
\caption{Time evolution of the density $n(x)$ and the phase $\theta (x, L_y/2)$ in the $x$ direction for the barrier velocities $v/v_c = 0.73$ (top panels) and $ 1.06$ (bottom panels).  
$n(x)$ is averaged over the initial ensemble and the $y$ direction, whereas $\theta (x, L_y/2)$ corresponds to a single sample. 
The dotted line denotes the barrier location. 
The blue triangle and red dot represent a vortex and an antivortex, respectively.   
}
\label{Fig:actime}
\end{figure}

\begin{figure}
\includegraphics[width=1.0\linewidth]{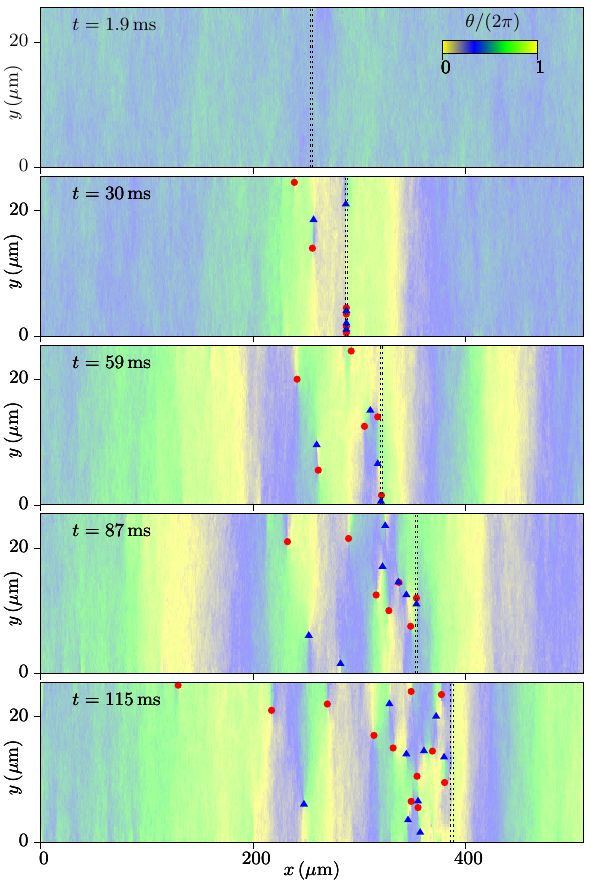}
\caption{Time evolution snapshots of the phase distribution $\theta (x, y)$ of a single sample for $v/v_c=1.06$. 
The barrier location at different times is marked by the two dashed vertical lines. 
The blue triangle and red dot represent a vortex and an antivortex, respectively.   
}
\label{Fig:acphase}
\end{figure}
\begin{figure}
\includegraphics[width=1.0\linewidth]{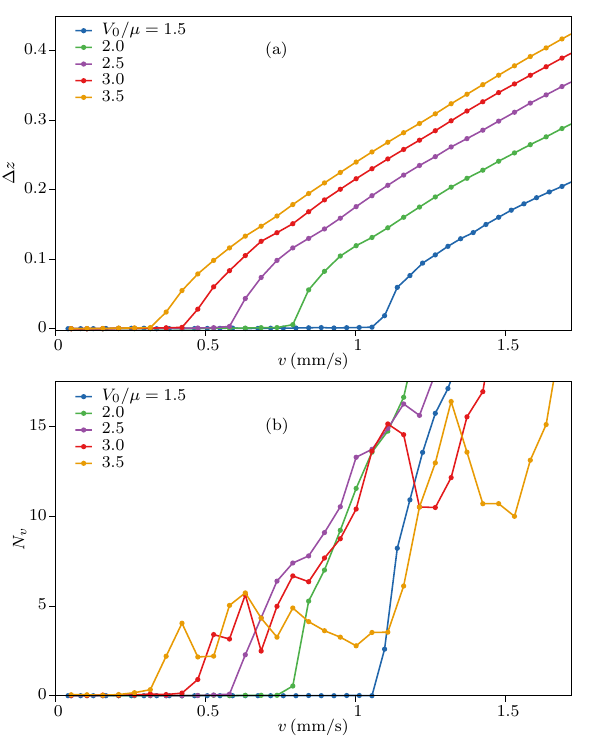}
\caption{(a) Imbalance $\Delta z$ as a function of the barrier velocity $v$ at varying barrier height $V_0$.  
(b) Total vortex number $N_v$ produced in the system for the same parameters as in (a). 
}
\label{Fig:acdis}
\end{figure}
\section{dc-ac transition in undriven junction}
\subsection{Estimate of the critical current}
Ref. \cite{Singh2020jj} derives an analytic estimate of the critical current by solving Eq. \ref{eq:GPE} and by considering single-particle tunneling across a rectangular barrier of height $V> \mu$ and width $d$, where $\mu$ is the mean-field energy. 
The critical current is given by $I_{c,\mp} = I_\mb t_0 (\tV, d)$, where $I_\mb$ is the bulk current and $t_0 (\tV, d)$ is the tunneling amplitude, with $\tV=V/\mu$. 
The variational solution gives \cite{Singh2020jj}
\begin{align}
t_0 (\tV, d) &= 2 \sqrt{2} \frac{\sqrt{6 \tV - 4 -  \sqrt{4 \tV^2 -2} }  } {2\tV + \sqrt{4 \tV^2 -2}  }\,  \nonumber \\
 & \quad \times  \exp \Bigl(- \frac{d}{2\xi} \sqrt{6 \tV - 4 -  \sqrt{4 \tV^2 -2} }     \Bigr).
\end{align}
We choose $\tV = \tV_0$ and $d \approx 1.1 w$ based on the values of $\tV_0$ and $w$ of the Gaussian barrier used in our simulations. 
The estimates of $I_{c,\mp}$ are compared with the simulation results in the main text.

\subsection{ac resistive dynamics}
In Fig. \ref{Fig:actime} we show the time evolution of the density $n(x)$ and the phase $\theta (x, L_y/2)$ along the $x$ direction for the barrier velocity $v$ below and above the critical velocity $v_c$. We average the density over the initial ensemble and the $y$ direction, 
whereas for the phase we use a single sample of the ensemble. 
As expected, for $v$ below $v_c$, the barrier does not create any strong perturbation of the density and the phase. 
Since we turn on $v$ instantaneously, there is the excitation of a weak-amplitude density pulse at time $t=0$, 
which damps out in long time evolution and does not contribute to any discernible imbalance change at time $t_f$.
For $v$ above $v_c$, the barrier motion induces a perturbation of the local density and phase near the barrier. 
During the evolution, the barrier creates the excitation of both vortex-antivortex pairs and phonons, 
which are observed as density waves.
This results in an overall density imbalance  between the left and right reservoir at time $t_f$. 
To understand the phase change in the $y$ direction and the mechanism of vortex nucleation 
we present the time evolution snapshots of the phase $\theta(x, y)$ in Fig. \ref{Fig:acphase}.  
During the evolution the vortex-antivortex pairs are nucleated from low density regions inside the barrier, 
which then propagate in the bulk behind the barrier. 
We identify vortex excitations by calculating the phase winding around the lattice plaquette of size $l\times l$ using $\sum_{\Box} \delta \theta(x,y) = \delta_x \theta(x,y) + \delta_y\theta(x+l,y)+\delta_x\theta(x+l,y+l)+\delta_y\theta(x,y+l)$,  
where $\theta(x, y)$ is the phase field of $\psi(x, y)$ and the phase differences between sites are taken to be $\delta_{x/y} \theta(x,y)  \in (-\pi, \pi]$. 
We associate a vortex (an antivortex) by a phase winding of $2\pi$ ($-2\pi$). 
By counting all vortices and antivortices we determine the total vortex number $N_\text{v}$ and average it over the initial ensemble.

In Fig. \ref{Fig:acdis}(a) we show the imbalance $\Delta z$ as a function of the barrier velocity $v$ at varying barrier height $V_0$. The onset of nonzero $\Delta z$ occurs at a lower $v$ for high $V_0$,  
marking the transition to the resistive regime of the junction. 
After the sharp onset $\Delta z$ features a linear dependence on $v$, which is clear at high $V_0$. 
This behavior is a manifestation of linear Ohmic dissipation, which occurs via the creation of vortex-antivortex pairs as shown in Fig.  \ref{Fig:acdis}(b).

\begin{figure}
\includegraphics[width=1.0\linewidth]{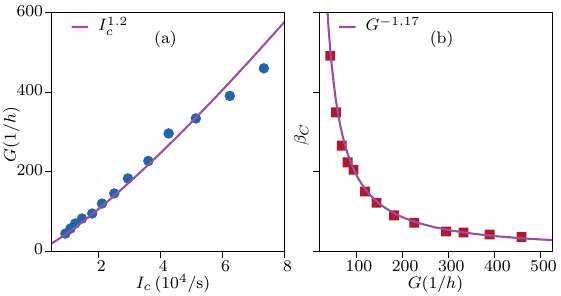}
\caption{(a) Conductance $G$ (dots) as  a function of $I_c$. The fit (continuous line) yields the dependence $G \propto I_c^{1.2}$. 
(b) Stewart-McCumber parameter $\beta_c$ (squares) as a function of $G$. 
We find the dependence $\beta_c \propto G^{-1.17}$ (continuous line). 
}
\label{Fig:scaling}
\end{figure}

\subsection{Scaling behavior of conductance and Stewart-McCumber parameter}
In this section we analyze the scaling behavior of  the conductance $G$ and the Stewart-McCumber parameter $\beta_c$, 
which are determined from the dc-ac transition of the junction for varying barrier height $\tV_0$.   
The dependence of $G$ on the critical current $I_c$ yields the scaling behavior $G \propto I_c^{1.2}$,  
which is shown in Fig. \ref{Fig:scaling}(a). 
This is close to the prediction of linear scaling based on the Ambegaokar-Baratoff relation of the critical current \cite{Ambegaokar1963}. 
Moreover, the measurements performed with molecular BEC report a quadratic scaling $G \propto I_c^{2}$ \cite{Kwon2020}, which deviates from our simulations. 

In Fig. \ref{Fig:scaling}(b) we plot the results of $\beta_c$ as a function of $G$,  
resulting in the dependence  $\beta_c \propto G^{-1.17}$.

\section{Dynamics of driven junction}
\begin{figure}
\includegraphics[width=1.0\linewidth]{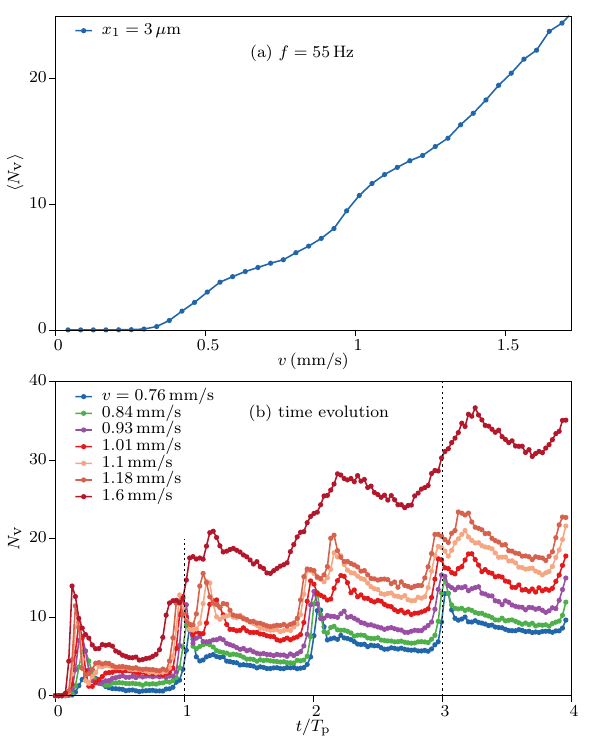}
\caption{Time-averaged vortex number $\langle N_\text{v} \rangle$ as a function of $v$ for the driving amplitude $x_1= 3\, \mum$ and frequency $f= 55\, \mHz$. 
(b) $N_\text{v} (t)$ as a function of $t/T_\text{p}$ for various values of the barrier velocity $v$, 
where $T_\text{p}=1/f$ is the time period. 
The two vertical dashed lines indicate the time window that we used for determining  the time-averaged response $\langle N_\text{v} \rangle$.  
}
\label{Fig:nvtime}
\end{figure}

\begin{figure}
\includegraphics[width=1.0\linewidth]{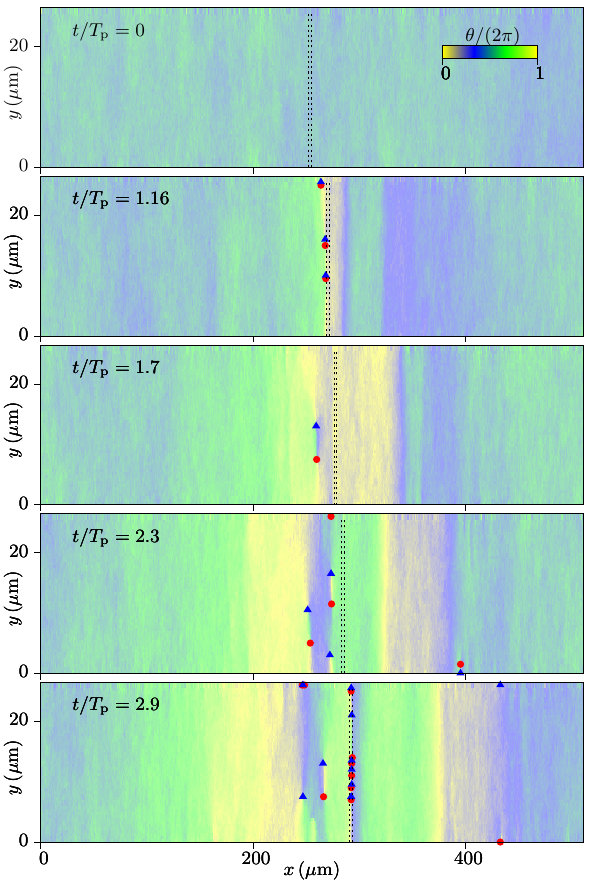}
\caption{Time evolution snapshots of the phase distribution $\theta (x, y)$ of a single sample for $v/v_c=0.6$,  $x_1= 3\, \mum$ and  $f= 55\, \mHz$. 
The barrier location at different times is marked by the two dashed vertical lines. 
The blue triangles and red dots represent vortices and antivortices, respectively.   
}
\label{Fig:dynamics}
\end{figure}

\subsection{Time-averaged vortex number}
In this section we describe how we obtain the time-averaged vortex number for the driven system. 
Fig. \ref{Fig:nvtime}(a) shows the results of time-averaged  vortex number $\langle N_\text{v} \rangle$ as a function of $v$, for $x_1= 3\, \mum$ and $f= 55\, \mHz$.  This is obtained in the following way. 
We first determine the time evolution of the total vortex number $N_\text{v}(t)$ during the driving, 
which is shown for varying barrier velocity $v$ in Fig. \ref{Fig:nvtime}(b). 
$N_\text{v}$ is calculated by counting all vortices and antivortices in the system and averaging it over the initial ensemble. 
$N_\text{v}(t)$ displays a peculiar behavior, showing vortex growths in periodic manner at times $t= n_c T_\mp$, 
where $n_c$ is the cycle number and $T_\mp = 1/f$ is the time period. 
This follows the argument of the barrier velocity $v(t)= v + 2\pi f x_1 \cos(2\pi f t)$ becoming maximum exactly at $t= n_c T_\mp$. 
At these times the barrier essentially surpasses the critical velocity and turn dissipative to loose energy via the creation of vortex-antivortex pairs. 
To obtain the time-averaged response we then average $N_\text{v}(t)$ over two cycles between $n_c=1$ and $n_c=3$, as indicated in Fig. \ref{Fig:nvtime}(b). 
This way we determine  $\langle N_\text{v} \rangle$ for various simulation parameters.

\subsection{Phase dynamics}
In Fig. \ref{Fig:dynamics} we show the dynamics of the phase $\theta (x, y)$ of a single sample at different times  
during the driving for $v/v_c=0.6$,  $x_1= 3\, \mum$ and  $f= 55\, \mHz$. 
The phase change across the barrier results in the creation of vortex-antivortex pairs, which then propagate in the bulk during the time evolution. 

\bibliography{References}